\documentstyle[]{article}
\newcommand{\bb}{\begin{eqnarray}}
\newcommand{\ee}{\end{eqnarray}}
\begin{document}
\title{T invariance of Higgs interactions in the standard model}

\author{P. Mitra\thanks{e-mail mitra@theory.saha.ernet.in}\\
Theory Group,\\ Saha Institute of Nuclear Physics,\\ Calcutta 700064}
\date{{\tt hep-ph/0303190}: presented at PASCOS'03}
\maketitle
\begin{abstract}
In the standard model, the Cabibbo-Kobayashi-Maskawa matrix, which
incorporates the time-reversal violation shown by the charged current weak
interactions, originates from the Higgs-quark interactions. 
The Yukawa interactions of quarks with the physical
Higgs particle can contain further complex phase factors, but 
nevertheless conserve T, as shown by constructing the fermion 
T transformation and the invariant euclidean fermion measure.
\end{abstract}

\section*{}

The quark mass terms arising from Yukawa interaction with Higgs fields
as a result of a Higgs vacuum expectation value are complex and of the form
$\bar q_L {M} q_R+\bar{\tilde q_L}{\tilde M}\tilde q_R+hc$
where $q,\tilde{q}$ stand for quarks with charge +2/3 and -1/3 respectively.
On diagonalization of ${M}, {\tilde M}$, by SU(3)$\times$SU(3) matrices
(without any U(1) factors to avoid QCD anomalies), 
\bb
q_L\rightarrow A_L^{-1}q_L,&\quad &q_R\rightarrow A_R^{-1}q_R\nonumber\\
\tilde q_L\rightarrow  \tilde A_L^{-1}\tilde q_L,
&\quad &\tilde q_R\rightarrow  \tilde A_R^{-1}\tilde q_R,
\ee
the W-interactions pick up a matrix $A_L \tilde A_L^{-1}\equiv {C}$, 
the Cabibbo-Kobayashi-Maskawa matrix, 
which may be complex and T-violating. 
The diagonalized mass terms may continue to be complex \cite{book}
and of the form $\bar\psi m\exp(i\theta'\gamma_5)\psi$.
The physical Higgs-quark interaction terms are proportional to these
mass terms and thus of the form $\bar\psi\phi\exp(i\theta'\gamma_5)\psi$.
The $\gamma_5$ phase factors are usually believed to violate both P and T. 

It is not widely known that these mass (and interaction) terms possess 
parity invariance \cite{bcm} with a {\it modified} quark transformation. 
Here we shall demonstrate time-reversal invariance.
A euclidean spacetime version of this invariance also exists.
This is necessary in setting up the euclidean functional integral.
A measure for quark functional integration can be constructed
to be invariant under this euclidean transformation in addition to
strong and electromagnetic gauge transformations.
This means that the time-reversal invariance has no QCD or even QED
anomaly. Any breaking of the invariance must be from other sources.

The requirement of time-reversal invariance of
a fermionic action amounts to demanding that
\bb
{\cal T}{\cal L}(t){\cal T}^{-1}
={\cal L}(-t).
\label{inv}\ee
Under the antilinear transformation ${\cal T}$, the gauge fields
$A_0(t) \rightarrow A_0(-t)$, $A_i(t)\rightarrow -A_i(-t)$, while
the fermion fields are taken to transform as
\bb
{\cal T}\psi(t){\cal T}^{-1}=T\psi(-t)
\ee
with suitable matrices $T$. If there exists a $T$ 
preserving (\ref{inv}), one has
time-reversal invariance, otherwise it is broken.

For real quark mass, we consider
${\cal L}=\bar\psi(i\gamma^\mu D_\mu-m)\psi$, where we keep in $D$ 
only vector gauge interactions with photons and gluons. The weak
interactions are understood to be treated separately in a perturbative
manner.
One finds that the matrices $T$ have to obey
\bb
T^\dagger T=1,\quad
-T^\dagger\gamma^{0*}\gamma^{i*}T=\gamma^0\gamma^i,\quad
T^\dagger\gamma^{0*}T=\gamma^0.
\ee
In the standard representation of gamma matrices, 
$\gamma^2$ is purely imaginary, while the rest are real. Then one finds 
$T=i\gamma^1\gamma^3$, the standard T-transformation for fermions.

For complex mass terms, the situation changes: the Lagrangian
${\cal L}=\bar\psi(i\gamma^\mu D_\mu-me^{i\theta'\gamma_5})\psi$
leads to the requirement
\bb
T^\dagger T=1,\quad
-T^\dagger\gamma^{0*}\gamma^{i*}T=\gamma^0\gamma^i,\quad
T^\dagger\gamma^{0*}e^{-i\theta'\gamma_5^*}T=\gamma^0e^{i\theta'\gamma_5}.
\ee
These are satisfied by 
$$T=ie^{i\theta'\gamma_5}\gamma^1\gamma^3.$$ 

To investigate the measure, one goes to the Euclidean metric. Then one has a
{\it linear} time-inversion instead of the antilinear time-reversal;
$A_0(t) \rightarrow -A_0(-t)$, $A_i(t)\rightarrow A_i(-t)$. Further,
$\psi,\bar\psi$ are independent fields, so that there are two
transformation matrices $T,\bar T$:
\bb
\psi(t)\rightarrow T\psi(-t),\quad
\bar\psi(t)\rightarrow \bar\psi(-t)\bar T.
\ee
$\gamma_0$ is also altered: all gamma matrices become antihermitian and
can be made hermitian by absorbing i. 
In this situation, the conditions for invariance are
\bb
-\bar T \gamma^0 T=\gamma^0,\quad
\bar T\gamma^{i}T=\gamma^i,\quad
\bar T T&=&1,
\ee
yielding
$\bar T=T=i\gamma^1\gamma^2\gamma^3$ when the mass term is real, and
\bb
-\bar T \gamma^0 T=\gamma^0,\quad
\bar T\gamma^{i}T=\gamma^i,\quad
\bar T e^{i\theta'\gamma_5}T=e^{i\theta'\gamma_5},
\ee
leading to
\bb
T=ie^{-i\theta'\gamma_5}\gamma^1\gamma^2\gamma^3,\quad
\bar T=ie^{i\theta'\gamma_5}\gamma^1\gamma^2\gamma^3,
\label{T}\ee
when the mass term is complex.

We come now to the construction of a  measure for fermion fields in 
euclidean spacetime and define the functional integral by
\bb
Z=\int {\cal D}A \prod_n \int da_n \prod_n \int d\bar{a}_n e^{-S}.
\ee
For a real quark mass term, $a_n,\bar{a}_n$ are defined by
\bb
\psi=\sum_n a_n\phi_n,\quad \bar\psi=\sum_n \bar{a}_n\phi^\dagger_n,
\ee
with $\phi_n$ being eigenfunctions of $i\gamma^\mu D_\mu$.

For a complex mass term,
it turns out to be necessary \cite{measure} to be more flexible 
and expand the quark fields as
\bb
\psi=e^{-i\beta\gamma^5/2}\sum_n a_n\phi_n,\quad \bar\psi=\sum_n 
\bar{a}_n\phi^\dagger_n e^{-i\beta\gamma^5/2},
\ee
with the phase $\beta$ to be determined. 
This is a $\beta$-dependent family of measures.
Note that $\theta',\beta$ can be removed
from the functional integral by transforming $a,\bar{a}$  
at the expense of changes in the $\theta F\tilde F$ term.
The effective parity and T violation parameter turns out to be
\bb
\bar\theta=\theta-\theta'+\beta.
\ee

$\beta$ is fixed by noting that
under the time-inversion operation for gauge fields,
\bb
\phi_n(x_0,\vec x)&\rightarrow&i\gamma^1\gamma^2\gamma^3\phi_n(-x_0,\vec x),\\
\phi_n^\dagger(x_0,\vec x)&\rightarrow&\phi_n^\dagger(-x_0,\vec x)i\gamma^1
\gamma^2\gamma^3,
\ee
so that
\bb
\bigg(e^{-{i\beta\gamma^5\over 2}}\phi_n(x_0,\vec x)\bigg)&\rightarrow&
ie^{-i\beta\gamma^5}\gamma^1\gamma^2\gamma^3
\bigg(e^{-{i\beta\gamma^5\over 2}}\phi_n(-x_0,\vec x)\bigg),\nonumber\\
\bigg(\phi_n^\dagger(x_0,\vec x)e^{-{i\beta\gamma^5\over 2}}\bigg)&\rightarrow&
\bigg(\phi_n^\dagger(-x_0,\vec x)e^{-{i\beta\gamma^5\over 2}}\bigg)
ie^{i\beta\gamma^5}\gamma^1\gamma^2\gamma^3.
\label{transf}\ee
The T-invariance of the measure or $a_n,\bar{a}_n$  
requires the consistency of the transformation matrices
in (\ref{transf}) with (\ref{T}). 
Like parity invariance \cite{measure}, this is achieved only with
\bb
\beta=\theta',
\ee
for which we have the interesting consequence \cite{measure} that
\bb 
\bar\theta=\theta -\theta'+\beta=\theta,
\ee
providing a resolution of the strong T or CP problem \cite{bcm}.

We have shown that even when the mass term is complex, it is
possible to define a time-reversal transformation whch keeps it
(as well as the kinetic terms and vector gauge interactions) 
invariant. This time-reversal is
not anomalous because it (or its euclidean version) can be preserved by
an appropriate quark measure. As regards the interaction
of the quark with the physical Higgs particle,
the phase in this Yukawa term 
is the same as the phase in the mass term which is generated
from the vacuum expectation value of the Higgs field. Consequently,
the time-reversal transformation defined in this paper keeps also
the quark-Higgs interaction term invariant. Time-reversal is of course
expected to be broken elsewhere: in the charged current weak interactions
through the CKM matrix and in the gluon sector through the vacuum angle
$\theta$ if it is nonzero. But the phase $\theta'$ in the quark mass term
left after diagonalization by SU(3)$\times$SU(3) matrices does not break T.

\end{document}